\documentclass[runningheads]{llncs}

\usepackage[table, dvipsnames]{xcolor}
\usepackage{eccv}

\usepackage{eccvabbrv}

\usepackage{graphicx}
\usepackage{booktabs}
\usepackage{comment}
\usepackage{lipsum} 
\usepackage{graphicx}
\usepackage{comment}
\usepackage{booktabs}
\usepackage{tabularx}
\usepackage{multirow}
\usepackage{xcolor,soul}
\usepackage{subcaption}
\usepackage{caption}
\usepackage{makecell}
\usepackage{adjustbox}

\usepackage{pifont}
\usepackage{xspace}

\definecolor{Gray}{gray}{0.9}
\definecolor{lgray}{gray}{0.95}
\definecolor{LightCyan}{rgb}{0.92,0.92,1}
\definecolor{lyellow}{rgb}{1,1,0.92}
\definecolor{lgreen}{rgb}{0.92,1,0.95}
\definecolor{llblue}{rgb}{0.92,0.93,0.95}
\definecolor{lred}{rgb}{1,0.85, 0.85}
\definecolor{tabhighlight}{HTML}{e5e5e5}

\usepackage{wrapfig}
\usepackage{subcaption}
\usepackage{colortbl}

\definecolor{cvprblue}{rgb}{0.21,0.49,0.74}
\definecolor{lblue}{rgb}{0.9,0.95,1}
\definecolor{lpurple}{rgb}{0.35,0.25,0.55}
\definecolor{lgreen}{rgb}{0.95,1,0.95}
\definecolor{sblue}{rgb}{0,0.45,1}
\definecolor{gold}{RGB}{248, 214, 99}
\definecolor{bronze}{RGB}{231, 188, 133}
\definecolor{silver}{RGB}{198, 210, 226}

\usepackage[accsupp]{axessibility}  

\usepackage[pagebackref,breaklinks,colorlinks,citecolor=eccvblue]{hyperref}

\usepackage{orcidlink}

\begin{document}

\title{Compressed Depth Map Super-Resolution and Restoration: AIM 2024 Challenge Results}

\titlerunning{AIM 2024 Compressed Depth Upsampling Challenge}

\author{
Marcos V. Conde\inst{1,2}$^{\dagger \ddagger}$\orcidlink{0000-0002-5823-4964} \and
Florin-Alexandru Vasluianu\inst{1}$^\dagger$\orcidlink{0009-0003-2366-8791}
\and
Jinhui Xiong\inst{3}$^\dagger$
\and
Wei Ye\inst{3}$^\dagger$
\and
Rakesh Ranjan\inst{3}$^\dagger$\orcidlink{0000-0003-3317-8473}
\and
Radu Timofte\inst{1}$^\dagger$\orcidlink{0000-0002-1478-0402}
\and \\
Huan Zheng \and Wencheng Han \and Tianyi Yan \and Jianbing Shen \and
Pihai Sun \and
Yuanqi Yao \and
Kui Jiang \and
Wenbo Zhao \and
Xianming Liu \and
Evgeny Burnaev \and
Junjun Jiang \and
Woojae Han \and Kyeonghyun Lee \and Seongmin Hong \and Se Young Chun \and
Jinseong Kim \and Dohyeong Kim \and Jeahwan Kim \and
Yubo Wang \and
Chi Zhang \and
Huizhen Luo \and
Yansai Wu \and
Mengcheng Huang \and
Chengji Liu \and
Chongli Yve \and
Jianhang Sun \and
Cheng Guo \and
Yingcai Du \and
Huang Jianhao \and Liu Shuai \and Li Chenghua
}

\authorrunning{Conde, Vasluianu, Xiong, Ye, Ranjan, Timofte, et al.}

\institute{
Computer Vision Lab, CAIDAS \& IFI, University of Würzburg \and
Visual Computing Group, FTG, Sony PlayStation \and
Meta Reality Labs\\
$^\dagger$ Challenge Organizers, $^\ddagger$ Corresponding Author \\
\url{https://ai4streaming-workshop.github.io/}
}

\maketitle


\vspace{-2.5mm}
\begin{figure}
    \centering
    \includegraphics[width=\linewidth]{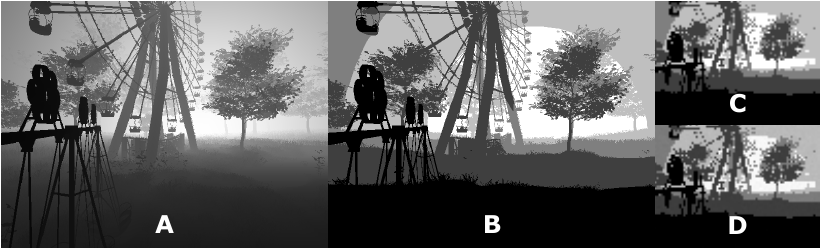}%
    \vspace{-2.3mm}
    \captionof{figure}{
    A graphical representation of the degradations suffered by a High-Resolution (HR) depth map (A), being mapped to its corresponding Low-Resolution (LR) version. Bitdepth reduction (B), spatial downscaling (C) and characteristic noise are applied to produce the Low-Quality (LQ) compressed depth map (D). 
    }
    \label{fig:teaser}
    \vspace{-7.5mm}
\end{figure}

\begin{abstract}
The increasing demand for augmented reality (AR) and virtual reality (VR) applications highlights the need for efficient depth information processing. Depth maps, essential for rendering realistic scenes and supporting advanced functionalities, are typically large and challenging to stream efficiently due to their size. This challenge introduces a focus on developing innovative depth upsampling techniques to reconstruct high-quality depth maps from compressed data. These techniques are crucial for overcoming the limitations posed by depth compression, which often degrades quality, loses scene details and introduces artifacts. By enhancing depth upsampling methods, this challenge aims to improve the efficiency and quality of depth map reconstruction. Our goal is to advance the state-of-the-art in depth processing technologies, thereby enhancing the overall user experience in AR and VR applications. 

\end{abstract}

\section{Introduction}
\label{sec:intro}
As the demand for immersive experiences in augmented reality (AR) and virtual reality (VR) applications continues to grow, efficient handling of depth information is crucial~\cite{krajancich2020optimizing, luo2020consistent}. Depth maps serve as an essential component for rendering realistic scenes and supporting various downstream applications such as object recognition~\cite{park2008multiple}, scene understanding ~\cite{krajancich2020optimizing}, and gesture tracking~\cite{yang2019gesture}. However, the high-resolution depth maps required for these applications are typically large in size, making them bandwidth-intensive and challenging to stream efficiently. Although real-time (and general) super-resolution of compressed images has been explored~\cite{zamfir2023towards, conde2023efficient, Conde_2024_CVPR, conde2022swin2sr}, there are not many depth upsampling techniques for recovering high-quality depth maps from compressed data.


Depth compression is necessary to enable the efficient transmission of depth information in real-time streaming applications, such as AR and VR. Compression reduces the amount of data that needs to be transmitted, thus reducing bandwidth requirements and latency. However, compression often introduces artifacts and reduces the quality of depth information. To mitigate these effects, depth upsampling techniques are employed to recover high-quality depth maps from downsampled and corrupted depth data.

Depth upsampling for compressed depth data is also related to the topic of depth completion~\cite{xu2019depth, hu2021penet} and depth densification~\cite{voynov2019perceptual, sun2023consistent}. Depth completion involves filling in missing depth information in incomplete depth maps. Depth densification targets to enhance the density of the depth map throughout. In many cases, depth sensors produce sparse depth maps with missing values due to occlusions, sensor noise, or limited sensor range and sensor resolution. Addressing these issues is crucial for tasks that require high-quality depth data, such as 3D reconstruction and detailed scene analysis.

The primary goal of the challenge is to encourage the development of novel depth upsampling techniques that achieve a balance between efficiency and depth map quality. By focusing on methods that are suitable for streaming environments, we aim to advance the state-of-the-art in real-time depth processing, ultimately enhancing the user experience in AR and VR applications.

\section{Challenge Dataset}
\label{sec:challenge}

The dataset used for the current edition of the Depth Compression through Super-Resolution and Refinement Challenge is based on the TartanAir data \cite{tartanair2020iros}. A subset of RGB images and depth maps corresponding to different scenes was sampled, being then used as the set of High-Quality (HQ) and High Resolution (HR) reference samples. Naturally, the subset of samples was split into two disjoint splits: 3866 for training, and 257 for testing the methods~\footnote{\url{https://codalab.lisn.upsaclay.fr/competitions/17339}}.

The splits do not share similar samples, 
but they are characterized by the same lossy compression. The participants never have access to the test ground-truth (HQ-HR depth maps).

\begin{figure}[t!]
    \centering
    \includegraphics[width=\linewidth]{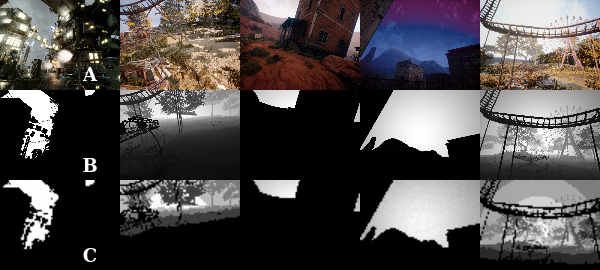}
    \caption{Samples from the Testing Phase split, consisting of the HR RGB image (A), the HR reference depth map (B), and the upscaled LR input depth map (C). The participants only have access to the HR RGB image and the LR Depth map.}
    \label{fig:dset_gallery}
\end{figure}

Consequently, we produce the set of Low-Quality (LQ) and Low-Resolution (LR) samples, corresponding to the HQ references, after suffering additional degradations -- see Figure \ref{fig:dset_gallery}. In the current challenge a bitdepth reduction operator is considered, with the 16-bit inputs being reduced to 12-bit. This operator is an irreversible transformation, with the lost local depth map smoothness posing a significant challenge in the depth map restoration process. Considering the $\downharpoonright$ operator as the bit resolution reduction operator, $	\downarrow_{8}$ the spatial $8\times$ downscaling operator, and $D$ the HQ depth map, we can define the LQ LR intermediary counterpart $d$ as follows:
\begin{equation}
    d = (D\downharpoonright)\downarrow_{8}
\end{equation}

For the final LQ sample $\hat{d}$, which is then used as an input sample in the restoration process, a characteristic noise profile is sampled, with the noise term $n$ then added to the intermediary LQ LR depth map $d$. The added noise corresponds to a combination of sensor reading noise and an Additive White Gaussian Noise (AWGN) component. Considering $\sigma_r^2$ the read noise variance and $\sigma_a^2$ the additive noise variance, the LQ depth map $\hat{d}$ is defined as follows:
\begin{equation}
\begin{split}
   n_r &\thicksim \mathcal{N}(0, \sigma_r^2) \\
   n_a &\thicksim \mathcal{N}(0, \sigma_a^2) \\
   n &= d \cdot n_r + n_a \\
   \hat{d} &= d + n
\end{split}
\end{equation}

In the current edition of the challenge the variance parameters were set to 0.05 for $\sigma_a^2$ and 0.02 for $\sigma_r^2$.


\paragraph{Associated AIM Challenges.} This challenge is one of the AIM 2024 Workshop\footnote{\url{https://www.cvlai.net/aim/2024/}} associated challenges on: sparse neural rendering~\cite{aim2024snr, aim2024snr_dataset}, UHD blind photo quality assessment~\cite{aim2024uhdbpqa}, compressed depth map super-resolution and restoration~\cite{aim2024cdmsrr}, efficient video super-resolution for AV1 compressed content~\cite{aim2024evsr}, video super-resolution quality assessment~\cite{aim2024vsrqa}, compressed video quality assessment~\cite{aim2024cvqa} and video saliency prediction~\cite{aim2024vsp}.

\section{Proposed Methods}
\label{sec:methods}

\begin{table*}[t]
    \centering
    \resizebox{\textwidth}{!}{
    \begin{tabular}{r c c c c c c}
        \toprule
        \rowcolor{lgray} Method & Ensemble & +Data & MAE~$\downarrow$ & RMSE~$\downarrow$ & \# Par. (M) & MACs (G) \\
        
        \midrule
        
         UM-IT~(\ref{sec:umit}) & Yes & No & \cellcolor{gold} \textbf{0.212} & \cellcolor{gold} \textbf{0.375} & 274.33  & 76.67 \\

         DAS-Depth~(\ref{sec:hit}) & No & Yes & \cellcolor{silver} 0.294 & \cellcolor{silver} 0.432 & 335.3 & 586.57 \\

         DINOv2 + ControlNet~(\ref{sec:icl}) & No & No & \cellcolor{bronze} 0.498 & \cellcolor{bronze} 0.816 & 52 & 483 \\

         RGA Inc.~(\ref{sec:rga}) & No & Yes & 0.512 & 1.621 & 41.4 & 121 \\

         RAFT-DU~(\ref{sec:custzs}) & No & No & 1.506 & 2.935 & 12 & 170 \\

         DAv2 ++~(\ref{sec:airia}) & Yes & No & 1.939 & 2.140 & 0.949 & 17.97 \\

         SGNet \cite{wang2024sgnet} & No & No & 1.337 & 1.854 & - & - \\

         Depth Anything V2~\cite{depth_anything_v2, depthanything} & No & No & 2.193 & 2.388 & - & - \\

         Bicubic Baseline & No & No & 16.48 & 57.69 & - & - \\
         

         
         \bottomrule
    \end{tabular}
    }
    \vspace{2mm}
    \caption{Compressed Depth Map Super-Resolution Challenge Results. We employ MAE and RMSE as the major metrics to select the best performing models. We highlight the top-3 (gold, silver, bronze) methods. SGNet \cite{wang2024sgnet} and Depth Anything V2~\cite{depth_anything_v2, depthanything} are baselines fine-tuned by the participants. "+Data" indicates if the solution uses additional data from the provided in the challenge. "\# Par." indicates the number of parameters in millions.
    }
    \vspace{-5mm}
    \label{tab:benchmark}
\end{table*}

\paragraph{\textbf{General Ideas}} Given an HR RGB image and the compressed LR depth map, the methods have to reconstruct the HR depth map.

In Table \ref{tab:benchmark} we provide the benchmark of the challenge, which includes computational complexity measured in MACs, and the number of parameters for each model. We can appreciate that simple Bicubic interpolation does not provide solid results as it usually happens in RGB upscaling, the reason is the lossy depth compression --- see Figure~\ref{fig:dset_gallery} (C) to appreciate the complex degradations.

\vspace{2mm}

Most state-of-the-art methods use the HR RGB image as the main source to predict the HR depth map. This is because the LR depth map, in some cases, due to the noise and compression does not contain meaningful information --- see Figure~\ref{fig:example_hit}. For instance, the top-3 methods: UM-IT~(\ref{sec:umit}), DAS-Depth~(\ref{sec:hit}) and DINOv2 + ControlNet~(\ref{sec:icl}), use encoder-decoder architectures to predict the HR depth map using mainly the HR RGB image, and the LR depth map as auxiliary information to condition the feaures. 

Many approaches rely on powerful pre-trained models such as DINO~\cite{oquab2024dinov2learningrobustvisual} and Depth Anything (DA)~\cite{depth_anything_v2, depthanything}. These methods are used as backbones, or for generating additional synthetic training data. 


\begin{figure*}[!ht]
    \centering
    \begin{subfigure}[t]{0.8\textwidth}
    \includegraphics[width=\textwidth]{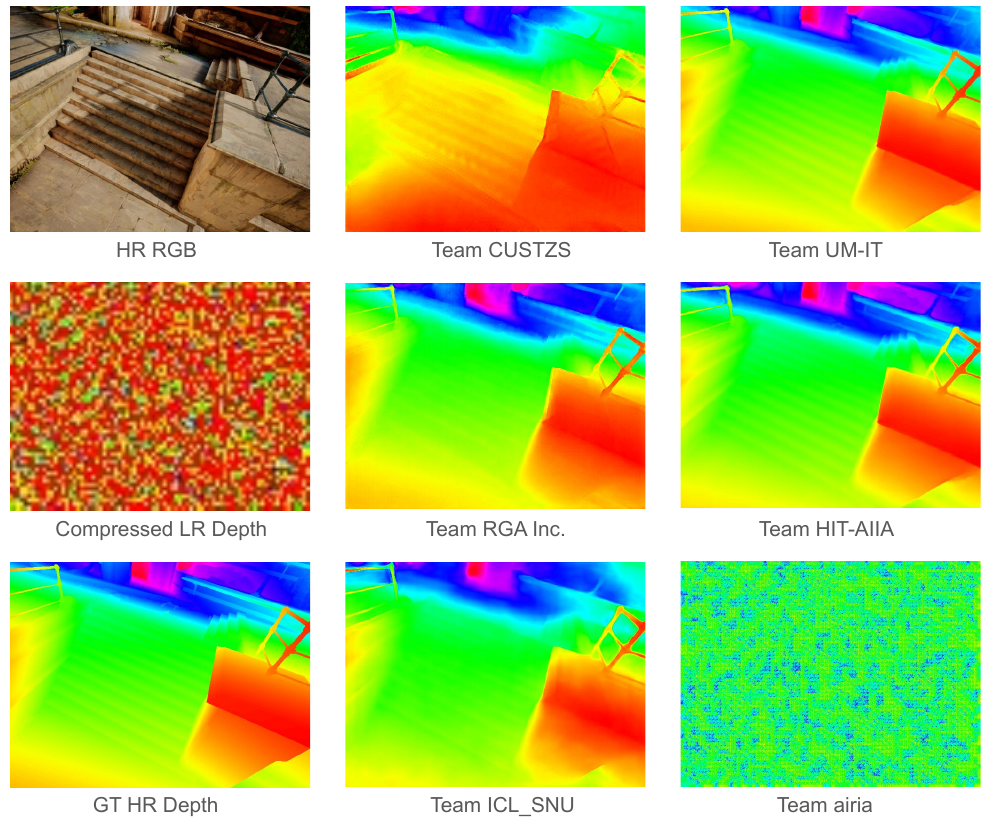}
    \end{subfigure}
    
    \begin{subfigure}[t]{0.8\textwidth}
    \includegraphics[width=\textwidth]{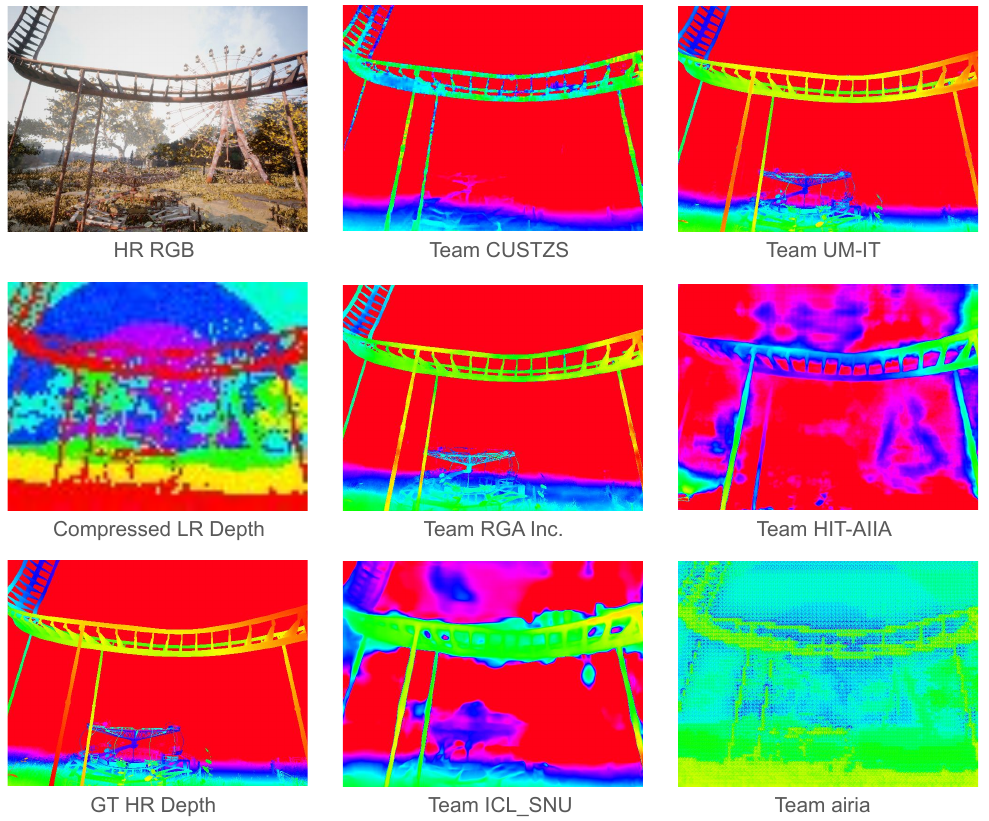}
    \end{subfigure}
    \caption{Visual comparisons of the proposed solutions on test samples.}
    \label{fig:visual_results}
\end{figure*} 

In the following sections, we describe the top solutions to the challenge. Please note that the method descriptions were provided by the respective teams or individual participants as their contributions to this report.

\newpage
\subsection{A Simple and Effective Baseline for Depth Upsampling and Refinement}
\label{sec:umit}

\emph{Huan Zheng, Wencheng Han, Tianyi Yan, Jianbing Shen} \\
\textit{SKL-IOTSC, CIS, University of Macau} (Team UM-IT)

\vspace{5mm}

\begin{figure*}[!ht]
    \centering
    \includegraphics[width=\linewidth]{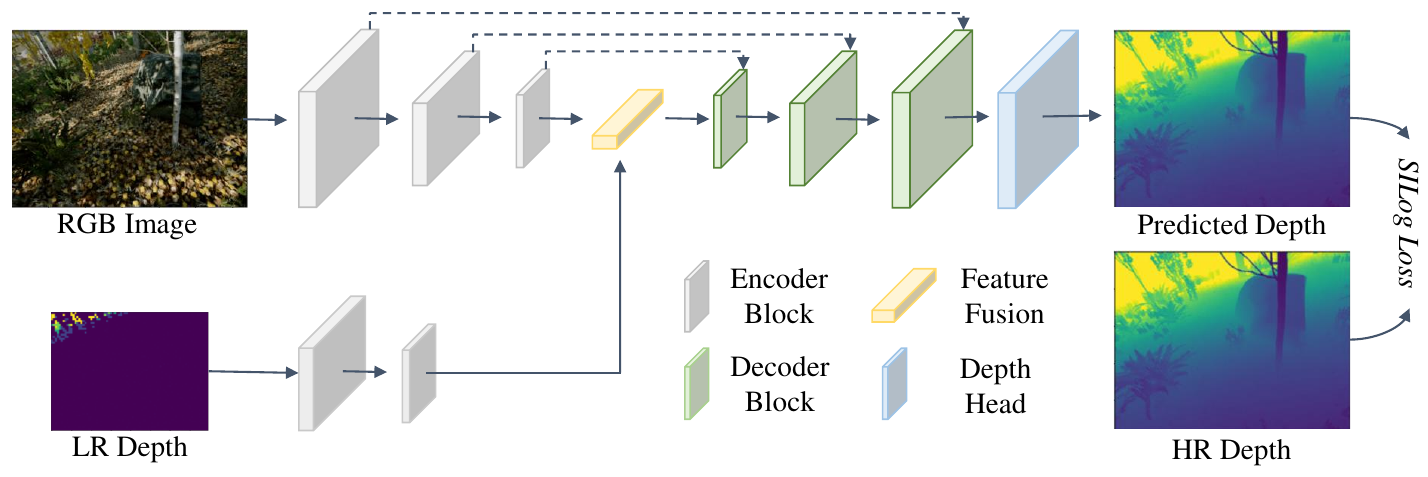}
    \caption{The overall framework of the team UM-IT proposal.}
    \label{fig:umit}
\end{figure*}

As shown in Fig. \ref{fig:umit}, we propose a simple and effective network for depth upsampling and refinement, utilizing a U-Net based architecture \cite{agarwal2023attention}. Specifically, our model takes one low-resolution depth map and one corresponding RGB image as inputs, aiming to produce a high-resolution depth map. For the input RGB image, a backbone network is employed to extract image features. In parallel, a low-resolution depth map is processed by a convolution-based sub-network for depth information extraction. The depth features and image features are then fused and fed into a transformer-based decoder. In the decoder, window-based cross-attention are applied, where the fused features and the image features from encoder are treated as query and key/value, respectively.
Finally, a depth prediction head generates the final depth output.

Furthermore, following previous work \cite{bhat2021adabins}, we use a scaled version of the Scale-Invariant loss (SILog) as the objective function for our method:
\begin{equation}
\mathcal{L}_{SILog} = \alpha \sqrt{\frac{1}{n} \sum_{i} g_{i}^{2} - \frac{\lambda}{n^{2}}\left(\sum_{i} g_{i}\right)^{2}},\
g_{i}=\log(\hat{d}_{i}) - \log(d_{i}),
\end{equation}

where $\hat{d}_{i}$ and $d_{i}$ denote the predicted depth and ground-truth depth at pixel $i$, respectively, and $n$ represents the total number of pixels in an image. Additionally, $\lambda$ and $\alpha$ are hyperparameters which are set to 0.85 and 10 in our experiments, respectively.

\subsubsection{Implementation details} 
The proposed method is implemented in PyTorch using 8 NVIDIA A100 GPUs. We use the Adam optimizer with a batch size of 16 and a weight decay of $1 \times 10^{-2}$. Our model is trained for 1500 epochs. For the first 400 epochs, the initial learning rate is set to $1 \times 10^{-4}$ and decreases linearly to $1 \times 10^{-5}$. For the remaining 1100 epochs, the learning rate is fixed at $1 \times 10^{-5}$.

During training, we employ various data augmentation techniques, including random horizontal flipping, random vertical flipping, and random cropping. Additionally, both the labeled depth and the low-resolution depth are clipped to a range of 0.1 to 20. We initialize the encoder of our model with the pretrained weights of Swin Transformer \cite{liu2021swin} on ImageNet \cite{deng2009imagenet}.

To enhance performance during inference, we use several strategies. First, the final depth values are obtained by averaging the predicted depths of the original image and its mirrored version. Second, the output depth values are clipped to the range of 0.1 to 20. Finally, we adopt an ensemble strategy that combines the depth predictions from several checkpoints to produce the final results.

We propose two versions of our model: a base version and a lite version. Compared to the base version, the lite version has fewer parameters (143.87M) and is more efficient (54.84G).

\subsection{DAS-Depth: Depth-Aware Scale Refinement for Monocular Depth Estimation}
\label{sec:hit}

\emph{Pihai Sun$^1$,
Yuanqi Yao$^1$,
Kui Jiang$^1$,
Wenbo Zhao$^1$,
Xianming Liu$^1$,
Evgeny Burnaev$^2$,
Junjun Jiang$^1$} \\
\textit{
$^1$ Harbin Institute of Technology (HIT)\\
$^2$ Skolkovo Institute of Science and Technology} (Team HIT-AIIA)

\vspace{5mm}

We explored the potential of zero-shot generalization using DepthAnything-V2~\cite{depth_anything_v2}, a state-of-the-art monocular depth estimation model. We employed the DepthAnything-V2-Large variant with its official pre-trained weights to generate disparity maps directly on the testing dataset. These disparity maps were then inverted to obtain the original depth maps. 

To align with the ground truth scale, we multiplied the depth values of the low-resolution depth map by a factor of 16 (We used the Theil-Sen regression method to fit the scale relationship between the mean of the ground truth depth maps in the training set and the mean of the corresponding low-resolution depth maps, resulting in a slope of approximately 16). Finally, we applied a least squares method to fit scaling and offset coefficients, ensuring proper alignment between the original depth map and the low-resolution depth map scales.

\begin{figure}[t] 
  \centering
  \includegraphics[width=\textwidth]{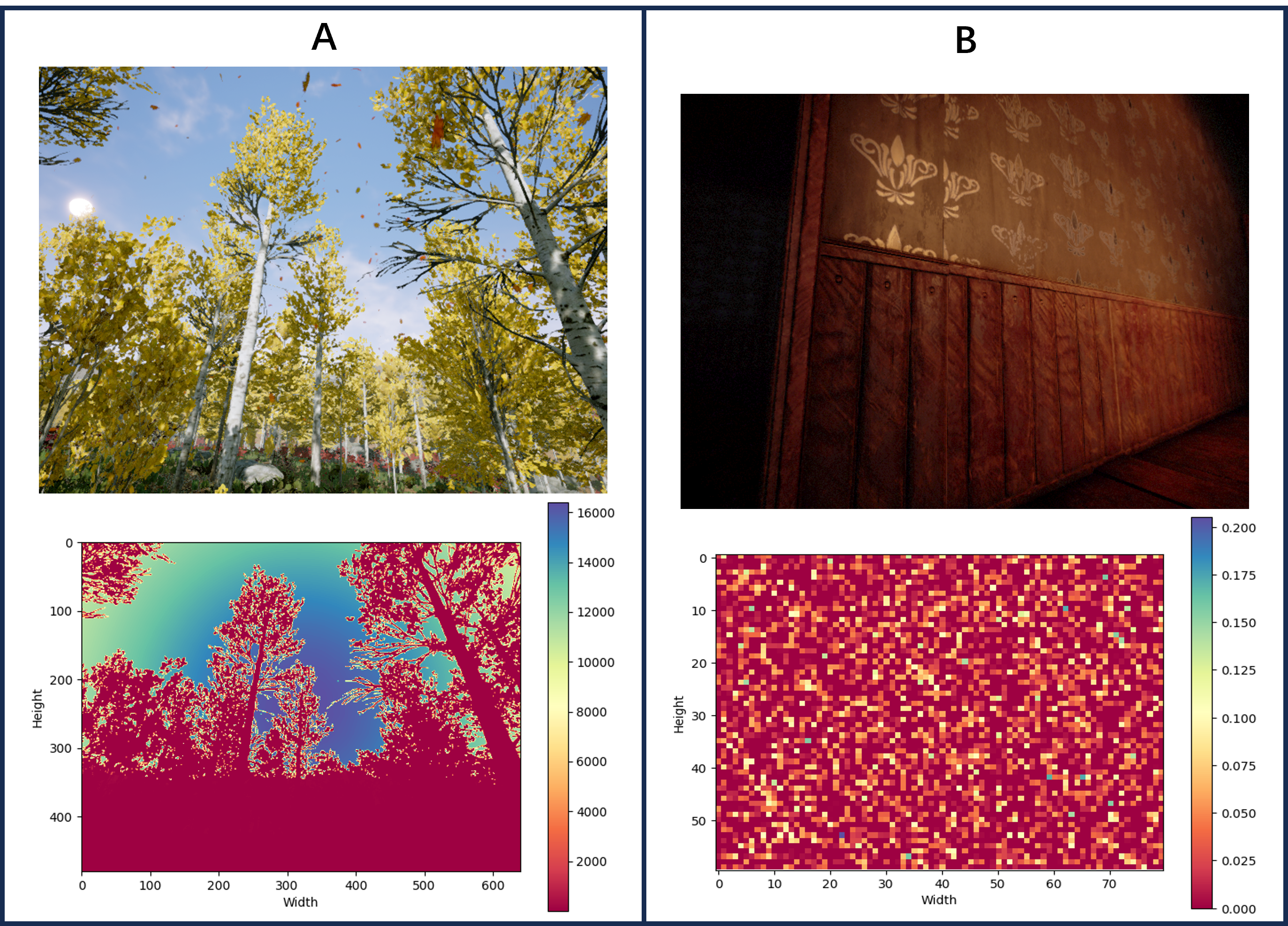} 
  \caption{DAS-Depth analysis of depth maps in the Training Set: (A) Inconsistent depth in Sky Regions; (B) Noise in Low-Resolution depth maps for smaller depth values}
  \label{fig:example_hit} 
\end{figure} 

\begin{figure*}[!ht]
    \centering
    \includegraphics[width=\textwidth]{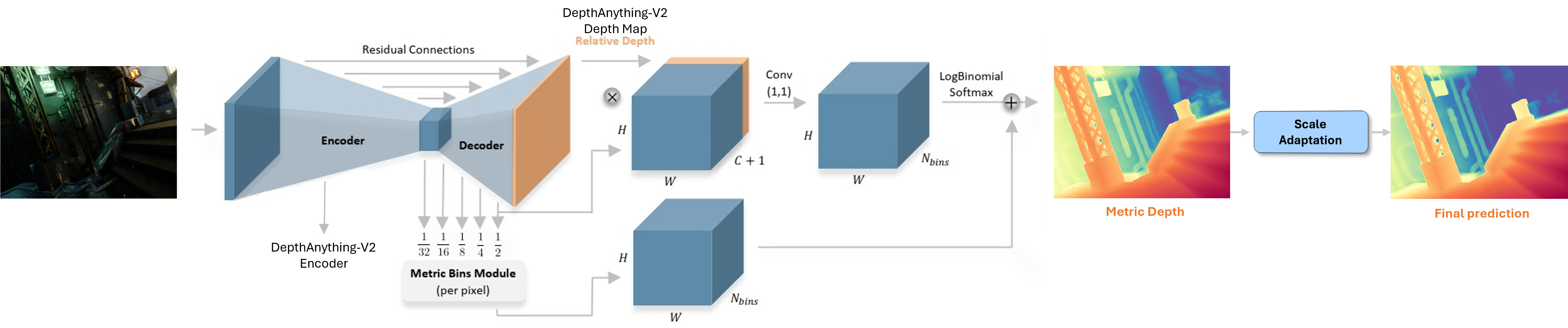}
    \caption{Overview of the DAS-Depth architecture proposed by team HIT-AIIA.}
    \label{fig:hit}
\end{figure*}

Based on this, we analyzed the challenge data and observed the following:

\begin{itemize}
\item The ground truth depth values have a large range and are unevenly distributed. The maximum depth in the ground truth provided by the training set reaches up to 65,504 meters, far exceeding the typical range required for depth estimation tasks. There are no pixels with depth values between 20,000 and 60,000. As shown in Figure~\ref{fig:example_hit} (A), pixels within the same sky region have different depth values. This indicates that the model's output space must be fully utilized.

\item As shown in Figure~\ref{fig:example_hit} (B), the low-resolution depth maps of samples with smaller depth values exhibit stronger noise. This means that using low-resolution depth maps as direct input to the model will significantly increase the difficulty of the model's learning process.
\end{itemize}

\noindent Based on the above observations, we chose to discard the low-resolution depth maps. By thoroughly analyzing the training set data, we constrained the model's output space to achieve optimal performance. \\

\noindent We initiated our training with the official public weights of Depth-Anything-V2-Large~\cite{depth_anything_v2}, a model renowned for its strong generalization capabilities due to its training on extensive synthetic and unlabeled real-world datasets. Our fine-tuning approach follows the ZoeDepth~\cite{bhat2023} pipeline but utilizes the pre-trained encoder from Depth-Anything-V2-Large instead of the MiDas encoder. \\

\noindent We constrain the output space of the model during training. During inference, we also adjust the scale of the predicted depth map to obtain the final depth map.

\subsubsection{Implementation details}

We utilized the Adam optimizer with learning rates set to 5e-6 for the encoder and 5e-5 for the decoder. The training was conducted on two RTX 3090 GPUs with a batch size of 8 for approximately 17 hrs. The training was carried out over a total of 100 epochs. We randomly divided 100 samples from the training set as the validation set. At the 96th epoch, the MAE indicator on the validation set achieved the best performance, and we selected this checkpoint as the final model weight.       

\subsection{Attaching ControlNet to an RGB2Depth Model for Depth Upsampling}
\label{sec:icl}

\emph{Woojae Han$^1$,
Kyeonghyun Lee$^1$,
Seongmin Hong$^1$,
Se Young Chun$^{1,2,3}$} \\
\textit{$^1$ Dept. of ECE, $^2$ INMC, $^3$ IPAI, (Team ICL\_SNU)\\ 
Seoul National University, Republic of Korea\\}

\vspace{5mm}

Our objective is to utilize DINOv2\cite{oquab2024dinov2learningrobustvisual} as the foundational model and make appropriate adjustments to enhance its performance. Specifically, we designed an approach to use low-resolution depth images as conditioning inputs, inspired by the concept introduced in ControlNet \cite{zhang2023addingconditionalcontroltexttoimage}. Our approach is structured around three major ideas: 

\begin{enumerate}
    \item Fine-tuning the model for optimal performance. 
    \item Data pre-processing to address the variability and noise in the dataset. 
    \item Implementing an effective loss function to train the model to capture detailed information while excluding background anomalies.
\end{enumerate}

\subsubsection{Method Description}

There are several well-trained models available for high resolution depth estimation from RGB images\cite{10.1145/3584860}. Among them, depth estimation model based on DINOv2 attached with depth prediction head (DPT head) is a good candidate. Cosidering DINOv2 as a backbone, we set our goal to fine-tune the DPT head to the given dataset. Since additional information of low resolution depth estimation is also given, we thought to bring the concept of ControlNet. We designed the basic output to be created with DINOv2 from RGB image while low resolution depth estimation is fed at the end of ouptut with appropriate zero convolution.

We only used the dataset given in the challenge without any augmentation or external datasets. Analyzing the data, we found that some of them containing outlier values in the background. In order to make our model to not fit only on background areas, we set a certain threshold and preprocess data to have threshold values if they are larger than it.

Another consideration of the dataset is the scale difference between low resolution and high resolution depth estimation. We found out that low resolution depths are not simply obtained with interpolation from high resolution depths. Due to this scale difference, we adopted SiLogLoss\cite{eigen2014depthmappredictionsingle} to effectively learn features within global scale difference. Eq.\ref{eq:silog} shows the loss function used in training process. $y$ is output of the model, $y^{*}$ is target high resolution depth estimation, and $\lambda$ is hyperparmeter of the loss. $y_{i}$, $y_{i}^{*}$ denotes each pixel value.

\begin{equation}\label{eq:silog}
L(y,y^{*}) = \frac{1}{n}\sum_{i}{(\log{y_{i}}-\log{y_{i}^{*}})^{2}} - \frac{\lambda}{n^{2}}\left( \sum_{i}(\log{y_{i}}-\log{y_{i}^{*}}) \right)^{2}
\end{equation}

Although use of fine tuned SGNet~\cite{wang2024sgnet} showed unsatisfactory results compared to our method, we have considered learning an ensemble of our method and SGNet to enhance boundary details of output depth estimation. However, ensemble of two models have shown bad results (MAE 0.6609, RMSE 1.0913) even with more training time. Although unsuccessful, the idea to consider boundaries with gradient mapping can be developed further to enhance our results.

\begin{figure*}[t]
    \centering
    \includegraphics[width=0.75\textwidth]{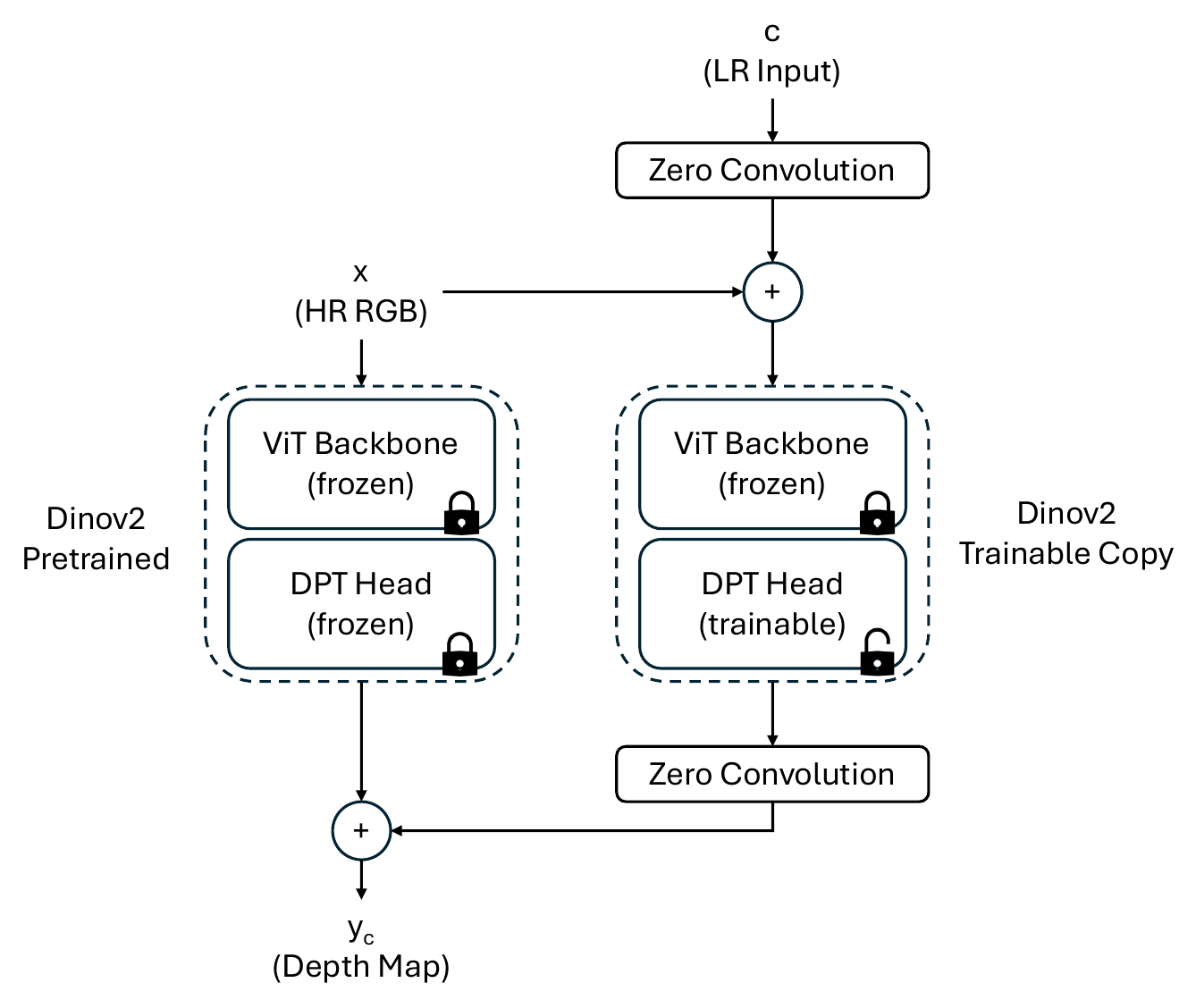}
    \caption{Model diagram of the ICL solution based on ControlNet \cite{zhang2023addingconditionalcontroltexttoimage}. Zero Convolution blocks and DPT Head in DINOv2 are fine-tuned.}
    \label{fig:icl}
\end{figure*}

\subsubsection{Implementation details}
All training sessions were conducted using one NVIDIA A100 GPU and pytorch environment. The Adam optimizer was employed for training. No external datasets were utilized; only the provided data was used for training and validation. Additionally, data augmentation techniques were not applied during the training process. The dataset was divided into training and validation sets with an 8:2 split to capture generalization ability. The training process was initiated with a learning rate of $10^{-4}$ since bigger values led to nan in loss and smaller values led to slow convergence of loss. Parameter $\lambda$ of SiLogLoss is set to $0.5$.

A specific pre-processing step was performed on the depth map data, where a threshold value of 25.0 was determined by manual visualization. This threshold was used to process the depth maps, ensuring that all pixel values exceeding this value were set to 25.0 to manage the background effectively. This pre-processing step helped in standardizing the input data and enhancing the model's focus on relevant features.

Our model was trained for a total of 48 epochs with the batch size set to 8. Each epoch consumed an average of 30 minutes. During training, the learning rate was set to 1e-4 for 42 epochs and changed to 1e-5 for the final 6 epochs.       

\subsection{Relative depth guided depth map upsampling}
\label{sec:rga}

\emph{Jinseong Kim,
Dohyeong Kim,
Jeahwan Kim} \\
\textit{RGA Inc., South Korea} (Team RGA Inc.)

\vspace{5mm}

\begin{figure*}[t]
    \centering
    \includegraphics[width=\textwidth]{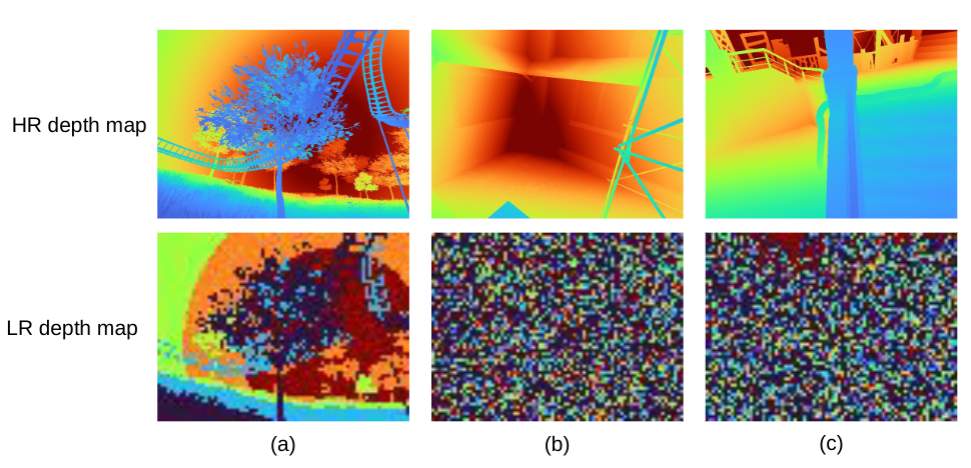}
    \caption{Examples of High-Resolution and Low-Resolution data pairs in the challenge dataset.}
    \label{fig:data_rga}
\end{figure*}

Before constructing the model, we analyzed the dataset used in the challenge. As seen in Fig~\ref{fig:data_rga}, when comparing the HR (High-Resolution) depth map with the LR (Low-Resolution) depth map, not only is the resolution reduced, but additional degradation is also observed. Moreover, some the LR depth maps are filled with noise, significantly reducing the reliability of depth information. 

To address the aforementioned issue, we utilize relative depth map as an additional guide. Using the DepthAnything~\cite{depthanything} model, we extract relative depth map and subsequently a Unet network fuses the relative depth map with a given LR depth map to reconstruct final depth map.

\begin{figure*}[!ht]
    \centering
    \includegraphics[width=\linewidth]{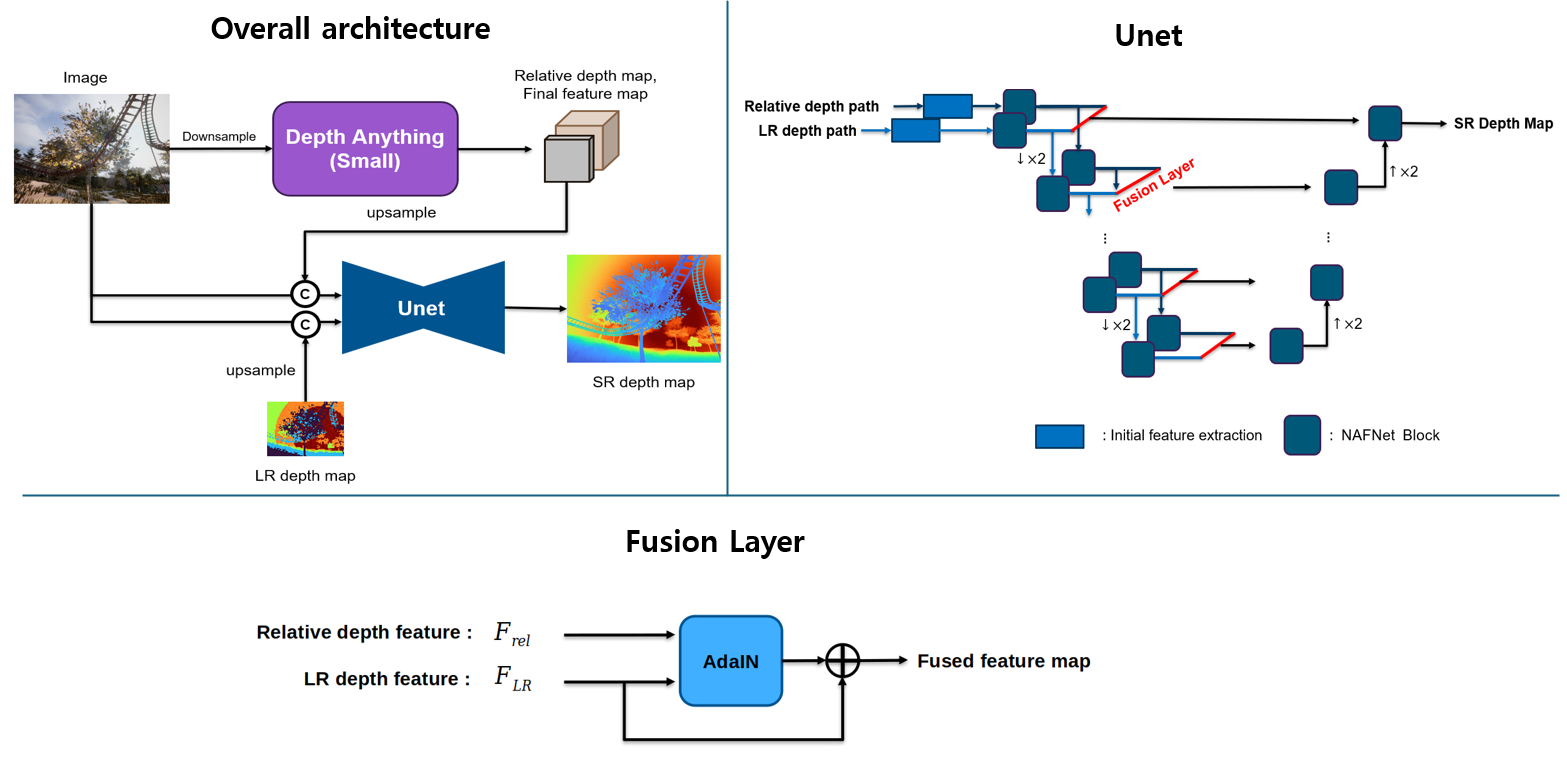}
    \caption{Overall architecture of the RGA network.}
    \label{fig:overall_rga}
\end{figure*}

\paragraph{\textbf{Model design.}}
Vit-S~\cite{vit} is used as the backbone network of DepthAnything~\cite{depthanything}. We utilize NAFNet~\cite{chu2022nafssr} blocks in the Unet network. AdaIN~\cite{adain} is used to fuse relative depth feature and LR depth feature. The solutions is illustrated in Figure~\ref{fig:overall_rga}.

Before training with the challenge dataset, we first trained our model on Dataset MVS-Synth~\cite{DeepMVS}. In depth map pre-processing, we clipped the depth maps from 0 to 300 and scaled them to a range of [0, 1]. We trained our model on the MVS-Synth dataset for 100 epochs. Afterward, we continued training on the challenge dataset until convergence. When training on the challenge dataset, the last 100 images of the training dataset were used for validation. We use $L1$ loss for network optimization.

\newpage
\subsection{RAFT-DU: Depth Upsampling Using RAFT Structures}
\label{sec:custzs}

\emph{Yubo Wang,
Chi Zhang,
Huizhen Luo,
Yansai Wu,
Mengcheng Huang,
Chengji Liu,
Chongli Yve,
Jianhang Sun,
Cheng Guo,
Yingcai Du} \\
\textit{CUSTZS, Zhongshan Research Institute of Changchun University of Science and Technology, China} (Team CUSTZS)

\vspace{5mm}

Our approach builds on RAFT-stereo\cite{lipson2021raft} by using the captured RGB image and the compressed depth as model inputs while recovering its modification from the original RAFT by extracting only the X-axis. We also found that directly calling the values of the original data helped a lot in testing the results compared to using normalized depth data. We also tried SGnet\cite{wang2024sgnet}, but this produces worse results.

As for the dataset we only used the dataset provided by the competition as for the adaptability and robustness of the scheme on other datasets we will do so after further attempts to improve the accuracy, we did not perform data augmentation and TTA nor did we use pre-trained models. This is likely to be our next step without increasing the overall computational cost. 

\begin{figure*}[t]
    \centering
    \includegraphics[width=1\textwidth]{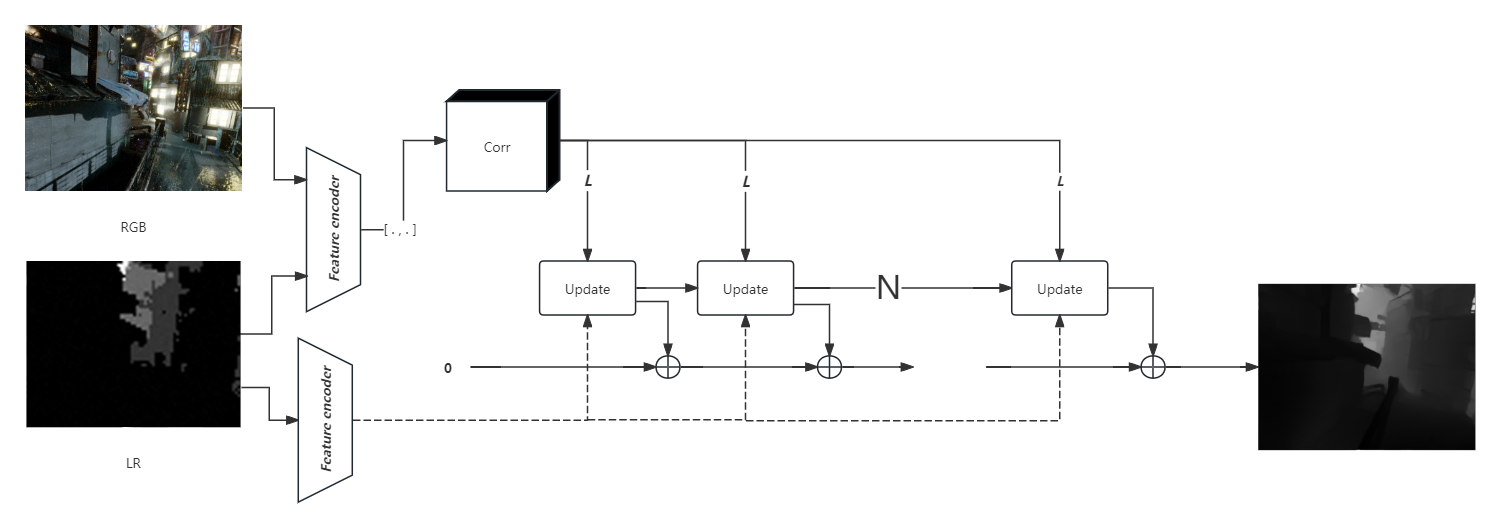}
    \caption{RAFT-DU overall structure.}
    \label{fig:custzs}
\end{figure*}


\paragraph{\textbf{Implementation details}}

Adam optimizer was used to train using an A100-40G at an initial learning rate of 0.0002, using only the training set provided by the competition, with a training time of roughly 30 hours. We use crops of size (256,256,3). No pre and post-processing or retraining of the data was performed, nor was any modification of the test data performed.    

\newpage
\begin{figure}[t]
    \centering
    \includegraphics[width=1\linewidth]{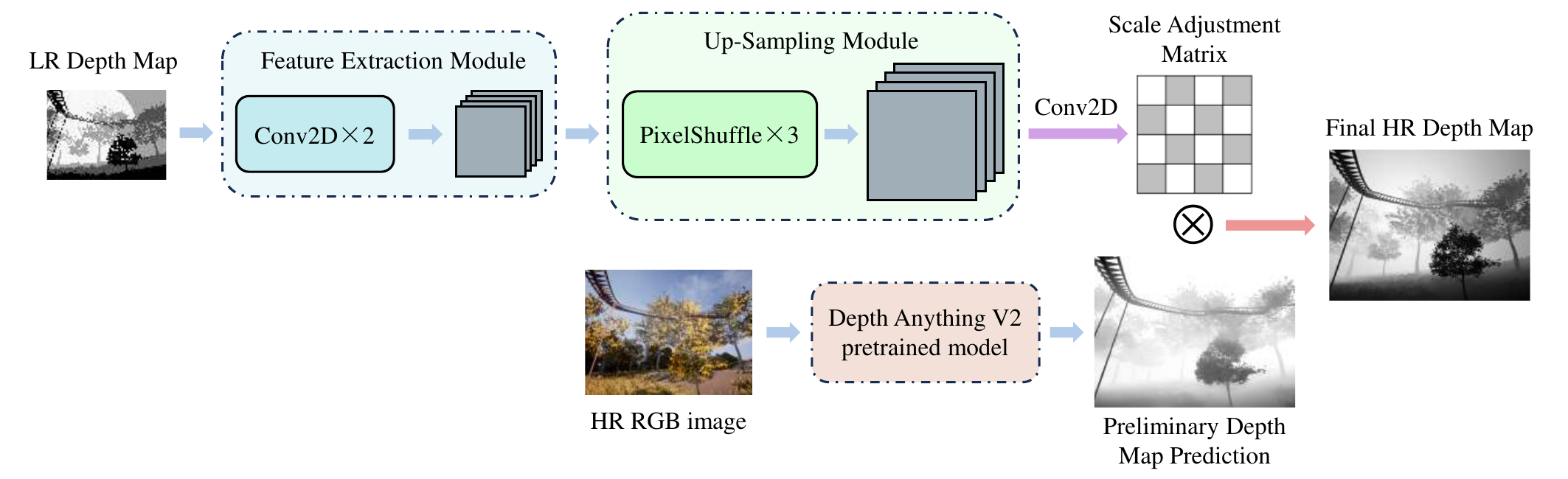}
    \caption{Team airia proposed Scale Adjustment Model.}
    \label{fig:my_diagram1}
\end{figure}

\subsection{A Fast Scale Adjustment Model for Depth Anything in Depth Up-sampling Challenge}
\label{sec:airia}

\emph{Jianhao Huang,
Shuai Liu,
Chenghua Li} \\
\textit{Team airia}

\vspace{5mm}

First, we directly use the training dataset of 3866 samples published on the Challenge website as our experimental dataset. Since only the training set contains ground truth, we divided one-tenth of the training set into verification sets to verify the performance of our model. In addition, we observed that the Depth Anything V2~\cite{depth_anything_v2} predictions had a smooth distribution of values, whereas the challenge provided a bumpy distribution of values in the data set. Therefore, for the Scale Adjustment model proposed by us, we choose to keep the value range of the low-resolution Depth map unchanged, and only implement the max-minimum normalization operation on the predicted results of Depth Anything V2~\cite{depth_anything_v2}. 

Before training begins, we first directly use Depth Anything V2 to load the "vitl" pre-training weights it provides, predicting a preliminary depth map for each RGB image in the training set. In the training process, our model first reads the low-resolution depth map and then gets a new two-dimensional array with the same shape as GT after processing by convolution and up-sampling modules. The model then reads the corresponding D.A. prediction depth map and performs element-wise multiplication with this two-dimensional array. Finally, the RMSE between the results obtained by multiplication and the GT depth map is used as a loss function to monitor the generation of the two-dimensional array.

During testing, we also first used Depth Anything V2 to predict a preliminary depth map for each RGB image in the test set. The Scale Adjustment model then reads a low-resolution depth map of each test data and generates a scale adjustment matrix using the trained model weights. Finally, the final depth map for each image will be the result of D.A.'s predicted depth map and element-wise multiplication of the scaling matrix.

\subsubsection{Model design} 

The overall framework of the Scale Adjustment model is shown in Fig~\ref{fig:my_diagram1}. In general, the model structure can be divided into two modules \textbf{feature extraction} and \textbf{up-sampling}. In the feature extraction module, the model uses two 2D convolution layers containing activation layers to extract features from low-resolution depth maps. Based on ensuring the concise and lightweight structure of the model, the potential scale information in the low-resolution depth map is identified as much as possible. In the up-sampling module, based on the work of Li et al.~\cite{Li_2018_ECCV}, the model uses three series \textbf{pixelshuffle layer} to up-sample the feature map. Each up-sampling layer doubles the height and width of the feature map. Finally, the model restores the channel number of the feature graph to 1 through a shape-invariant 2D convolution layer and obtains a scale adjustment vector of the shape (1, H, W).



\subsubsection{Implementation details}

In the experiment, our model trained a total of 500 epochs with Adam as the optimizer, 0.001 as the initial learning rate, and 128 as the batch size. All training and testing procedures are performed on an NVIDIA A100 GPU on a remote server. 



\section{Conclusion}
In this paper, we presented the results of the compressed depth map super-resolution challenge held alongside ECCV 2024. The submissions featured a variety of approaches, with some demonstrating superior performance. Notably, methods utilizing pre-trained large backbone networks effectively preserved image features and transferred them to the reconstructed depth features (as illustrated in fig.~\ref{fig:visual_results}), indicating that pre-trained models on image domains can be adapted to depth domains with effective fine-tuning. Looking ahead, a key focus can be on reducing model complexity to enable real-time applications and facilitate deployment on portable devices.

\section*{Acknowledgements}
This work was partially supported by the Humboldt Foundation. We thank the AIM 2024 sponsors: Meta Reality Labs, KuaiShou, Huawei, Sony Interactive Entertainment, and the University of W\"urzburg (Computer Vision Lab).



\bibliographystyle{splncs04}
\bibliography{main}

\begin{thebibliography}{10}
\providecommand{\url}[1]{\texttt{#1}}
\providecommand{\urlprefix}{URL }
\providecommand{\doi}[1]{https://doi.org/#1}

\bibitem{agarwal2023attention}
Agarwal, A., Arora, C.: Attention attention everywhere: Monocular depth prediction with skip attention. In: Proceedings of the IEEE/CVF Winter Conference on Applications of Computer Vision. pp. 5861--5870 (2023)

\bibitem{bhat2021adabins}
Bhat, S.F., Alhashim, I., Wonka, P.: Adabins: Depth estimation using adaptive bins. In: Proceedings of the IEEE/CVF conference on computer vision and pattern recognition. pp. 4009--4018 (2021)

\bibitem{bhat2023}
Bhat, S.F., Birkl, R., Wofk, D., Wonka, P., Müller, M.: Zoedepth: Zero-shot transfer by combining relative and metric depth (2023). \doi{10.48550/ARXIV.2302.12288}, \url{https://arxiv.org/abs/2302.12288}

\bibitem{chu2022nafssr}
Chu, X., Chen, L., Yu, W.: Nafssr: Stereo image super-resolution using nafnet. In: IEEE/CVF Conference on Computer Vision and Pattern Recognition (CVPR) Workshops (2022)

\bibitem{conde2022swin2sr}
Conde, M.V., Choi, U.J., Burchi, M., Timofte, R.: Swin2sr: Swinv2 transformer for compressed image super-resolution and restoration. In: European Conference on Computer Vision. pp. 669--687. Springer (2022)

\bibitem{aim2024evsr}
Conde, M.V., Lei, Z., Li, W., Bampis, C., Katsavounidis, I., Timofte, R., et~al.: {{AIM} 2024 Challenge on Efficient Video Super-Resolution for AV1 Compressed Content}. In: Proceedings of the European Conference on Computer Vision (ECCV) Workshops (2024)

\bibitem{Conde_2024_CVPR}
Conde, M.V., Lei, Z., Li, W., Katsavounidis, I., Timofte, R., Yan, M., Liu, X., Wang, Q., Ye, X., Du, Z., Zhang, T., Li, Z., Wei, H., Ge, C., Lv, J., Sun, L., Pan, J., Dong, J., Tang, J., Zhou, M., Yan, Y., Yoon, K., Gankhuyag, G., Lee, J.H., Choi, U.J., Moon, H.C., Jeong, T.H., Yang, Y., Kim, J.G., Jeong, J., Kim, S., Qiu, X., Zhou, Y., Wu, K., Dai, X., Tang, H., Deng, W., Gao, Q., Tong, T., Peng, L., Guo, J., Di, X., Liao, B., Du, Z., Xia, P., Pei, R., Wang, Y., Cao, Y., Zha, Z., Han, B., Yu, H., Wu, Z., Wan, C., Liu, Y., Yu, H., Li, J., Huang, Z., Huang, Y., Zou, Y., Guan, X., Jia, Q., Zhang, H., Yin, X., Zuo, K., Zhang, D., Liu, T., Chen, H., Jin, Y.: Real-time 4k super-resolution of compressed avif images. ais 2024 challenge survey. In: Proceedings of the IEEE/CVF Conference on Computer Vision and Pattern Recognition (CVPR) Workshops. pp. 5838--5856 (June 2024)

\bibitem{aim2024cdmsrr}
Conde, M.V., Vasluianu, F.A., Xiong, J., Ye, W., Ranjan, R., Timofte, R., et~al.: {Compressed Depth Map Super-Resolution and Restoration: {AIM} 2024 Challenge Results}. In: Proceedings of the European Conference on Computer Vision (ECCV) Workshops (2024)

\bibitem{conde2023efficient}
Conde, M.V., Zamfir, E., Timofte, R., Motilla, D., Liu, C., Zhang, Z., Peng, Y., Lin, Y., Guo, J., Zou, X., et~al.: Efficient deep models for real-time 4k image super-resolution. ntire 2023 benchmark and report. In: Proceedings of the IEEE/CVF conference on computer vision and pattern recognition. pp. 1495--1521 (2023)

\bibitem{deng2009imagenet}
Deng, J., Dong, W., Socher, R., Li, L.J., Li, K., Fei-Fei, L.: Imagenet: A large-scale hierarchical image database. In: 2009 IEEE conference on computer vision and pattern recognition. pp. 248--255. Ieee (2009)

\bibitem{vit}
Dosovitskiy, A., Beyer, L., Kolesnikov, A., Weissenborn, D., Zhai, X., Unterthiner, T., Dehghani, M., Minderer, M., Heigold, G., Gelly, S., Uszkoreit, J., Houlsby, N.: An image is worth 16x16 words: Transformers for image recognition at scale (2021), \url{https://arxiv.org/abs/2010.11929}

\bibitem{eigen2014depthmappredictionsingle}
Eigen, D., Puhrsch, C., Fergus, R.: Depth map prediction from a single image using a multi-scale deep network (2014), \url{https://arxiv.org/abs/1406.2283}

\bibitem{aim2024uhdbpqa}
Hosu, V., Conde, M.V., Agnolucci, L., Barman, N., Zadtootaghaj, S., Timofte, R., et~al.: {AIM} 2024 challenge on uhd blind photo quality assessment. In: Proceedings of the European Conference on Computer Vision (ECCV) Workshops (2024)

\bibitem{hu2021penet}
Hu, M., Wang, S., Li, B., Ning, S., Fan, L., Gong, X.: Penet: Towards precise and efficient image guided depth completion. In: 2021 IEEE International Conference on Robotics and Automation (ICRA). pp. 13656--13662. IEEE (2021)

\bibitem{DeepMVS}
Huang, P.H., Matzen, K., Kopf, J., Ahuja, N., Huang, J.B.: Deepmvs: Learning multi-view stereopsis. In: IEEE/CVF Conference on Computer Vision and Pattern Recognition (CVPR) (2018)

\bibitem{adain}
Huang, X., Belongie, S.: Arbitrary style transfer in real-time with adaptive instance normalization. In: IEEE International Conference on Computer Vision (ICCV) (2017)

\bibitem{krajancich2020optimizing}
Krajancich, B., Kellnhofer, P., Wetzstein, G.: Optimizing depth perception in virtual and augmented reality through gaze-contingent stereo rendering. ACM Transactions on Graphics (TOG)  \textbf{39}(6),  1--10 (2020)

\bibitem{Li_2018_ECCV}
Li, J., Fang, F., Mei, K., Zhang, G.: Multi-scale residual network for image super-resolution. In: Proceedings of the European Conference on Computer Vision (ECCV) (September 2018)

\bibitem{lipson2021raft}
Lipson, L., Teed, Z., Deng, J.: Raft-stereo: Multilevel recurrent field transforms for stereo matching. In: 2021 International Conference on 3D Vision (3DV). pp. 218--227. IEEE (2021)

\bibitem{liu2021swin}
Liu, Z., Lin, Y., Cao, Y., Hu, H., Wei, Y., Zhang, Z., Lin, S., Guo, B.: Swin transformer: Hierarchical vision transformer using shifted windows. In: Proceedings of the IEEE/CVF international conference on computer vision. pp. 10012--10022 (2021)

\bibitem{luo2020consistent}
Luo, X., Huang, J.B., Szeliski, R., Matzen, K., Kopf, J.: Consistent video depth estimation. ACM Transactions on Graphics (ToG)  \textbf{39}(4),  71--1 (2020)

\bibitem{aim2024vsrqa}
Molodetskikh, I., Borisov, A., Vatolin, D.S., Timofte, R., et~al.: {{AIM} 2024 Challenge on Video Super-Resolution Quality Assessment: Methods and Results}. In: Proceedings of the European Conference on Computer Vision (ECCV) Workshops (2024)

\bibitem{aim2024vsp}
Moskalenko, A., Bryntsev, A., Vatolin, D.S., Timofte, R., et~al.: {AIM} 2024 challenge on video saliency prediction: Methods and results. In: Proceedings of the European Conference on Computer Vision (ECCV) Workshops (2024)

\bibitem{aim2024snr}
Nazarczuk, M., Catley-Chandar, S., Tanay, T., Shaw, R., Pérez-Pellitero, E., Timofte, R., et~al.: {{AIM} 2024 Sparse Neural Rendering Challenge: Methods and Results}. In: Proceedings of the European Conference on Computer Vision (ECCV) Workshops (2024)

\bibitem{aim2024snr_dataset}
Nazarczuk, M., Tanay, T., Catley-Chandar, S., Shaw, R., Timofte, R., Pérez-Pellitero, E.: {{AIM} 2024 Sparse Neural Rendering Challenge: Dataset and Benchmark}. In: Proceedings of the European Conference on Computer Vision (ECCV) Workshops (2024)

\bibitem{oquab2024dinov2learningrobustvisual}
Oquab, M., Darcet, T., Moutakanni, T., Vo, H., Szafraniec, M., Khalidov, V., Fernandez, P., Haziza, D., Massa, F., El-Nouby, A., Assran, M., Ballas, N., Galuba, W., Howes, R., Huang, P.Y., Li, S.W., Misra, I., Rabbat, M., Sharma, V., Synnaeve, G., Xu, H., Jegou, H., Mairal, J., Labatut, P., Joulin, A., Bojanowski, P.: Dinov2: Learning robust visual features without supervision (2024), \url{https://arxiv.org/abs/2304.07193}

\bibitem{park2008multiple}
Park, Y., Lepetit, V., Woo, W.: Multiple 3d object tracking for augmented reality. In: 2008 7th IEEE/ACM International Symposium on Mixed and Augmented Reality. pp. 117--120. IEEE (2008)

\bibitem{aim2024cvqa}
Smirnov, M., Gushchin, A., Antsiferova, A., Vatolin, D.S., Timofte, R., et~al.: {{AIM} 2024 Challenge on Compressed Video Quality Assessment: Methods and Results}. In: Proceedings of the European Conference on Computer Vision (ECCV) Workshops (2024)

\bibitem{sun2023consistent}
Sun, Z., Ye, W., Xiong, J., Choe, G., Wang, J., Su, S., Ranjan, R.: Consistent direct time-of-flight video depth super-resolution. In: Proceedings of the ieee/cvf conference on computer vision and pattern recognition. pp. 5075--5085 (2023)

\bibitem{voynov2019perceptual}
Voynov, O., Artemov, A., Egiazarian, V., Notchenko, A., Bobrovskikh, G., Burnaev, E., Zorin, D.: Perceptual deep depth super-resolution. In: Proceedings of the ieee/cvf international conference on computer vision. pp. 5653--5663 (2019)

\bibitem{tartanair2020iros}
Wang, W., Zhu, D., Wang, X., Hu, Y., Qiu, Y., Wang, C., Hu, Y., Kapoor, A., Scherer, S.: Tartanair: A dataset to push the limits of visual slam  (2020)

\bibitem{wang2024sgnet}
Wang, Z., Yan, Z., Yang, J.: Sgnet: Structure guided network via gradient-frequency awareness for depth map super-resolution. In: Proceedings of the AAAI Conference on Artificial Intelligence. vol.~38, pp. 5823--5831 (2024)

\bibitem{xu2019depth}
Xu, Y., Zhu, X., Shi, J., Zhang, G., Bao, H., Li, H.: Depth completion from sparse lidar data with depth-normal constraints. In: Proceedings of the IEEE/CVF International Conference on Computer Vision. pp. 2811--2820 (2019)

\bibitem{yang2019gesture}
Yang, L., Huang, J., Feng, T., Hong-An, W., Guo-Zhong, D.: Gesture interaction in virtual reality. Virtual Reality \& Intelligent Hardware  \textbf{1}(1),  84--112 (2019)

\bibitem{depthanything}
Yang, L., Kang, B., Huang, Z., Xu, X., Feng, J., Zhao, H.: Depth anything: Unleashing the power of large-scale unlabeled data. In: IEEE/CVF Conference on Computer Vision and Pattern Recognition (CVPR) (2024)

\bibitem{depth_anything_v2}
Yang, L., Kang, B., Huang, Z., Zhao, Z., Xu, X., Feng, J., Zhao, H.: Depth anything v2. arXiv preprint arXiv:2406.09414  (2024)

\bibitem{zamfir2023towards}
Zamfir, E., Conde, M.V., Timofte, R.: Towards real-time 4k image super-resolution. In: Proceedings of the IEEE/CVF Conference on Computer Vision and Pattern Recognition. pp. 1522--1532 (2023)

\bibitem{zhang2023addingconditionalcontroltexttoimage}
Zhang, L., Rao, A., Agrawala, M.: Adding conditional control to text-to-image diffusion models (2023), \url{https://arxiv.org/abs/2302.05543}

\bibitem{10.1145/3584860}
Zhong, Z., Liu, X., Jiang, J., Zhao, D., Ji, X.: Guided depth map super-resolution: A survey. ACM Comput. Surv.  \textbf{55}(14s) (jul 2023). \doi{10.1145/3584860}, \url{https://doi.org/10.1145/3584860}

\end{thebibliography}

\end{document}